# Phonon-drag studies of (110) AlAs quantum wells

D Lehmann[1] and C Jasiukiewicz[2]

[1] Institute of Theoretical Physics, Technische Universität Dresden, 01062 Dresden, Germany
[2] Faculty of Mathematics and Applied Physics, Rzeszow University of Technology, ul. W. Pola 2, 35-959 Rzeszow, Poland

E-mail: Dietmar.G.Lehmann@tu-dresden.de

**Abstract**. We investigate the possibility of using phonon-drag imaging for the study of 2D electrons in (110) AlAs quantum wells. Our numerical simulations show that direct information on the strain and quantum confinement dependent valley occupancy of the electrons, on the anisotropic effective mass tensor and on electron-phonon coupling parameters can be obtained.

## 1. Introduction

Quasi two-dimensional (2D) electron systems in AlAs quantum wells (QWs) have shown a variety of novel phenomena (for a review see [1]). AlAs, though closely related to GaAs in structure, differs significantly from GaAs in its electronic properties. The electrons confined to an AlAs layer occupy multiple conduction-band minima at the X-points of the Brillouin zone. Compared to the standard quasi-2D electron system in a GaAs QW or heterojunction, where the electrons occupy the nondegenerate (excluding spin) conduction band minimum at the Γ-point of the Brillouin zone, the electrons in AlAs QWs possess a much larger and anisotropic effective mass, as well as a much larger (and of different sign) effective g-factor. Additionally, valley degeneracy occurs, which can be tuned by quantum confinement and strain. Due to the larger effective mass not only the 2D electron system is more interacting (the ratio of Coulomb to kinetic energy increases), we have also a stronger electron-phonon interaction. Furthermore, due to the anisotropy in the matrix elements of deformation potential electron-phonon scattering, a more effective coupling of the transverse acoustic phonons in AlAs QWs, compared to GaAs, occurs.

Phonon spectroscopy techniques, like phonon-drag imaging, are well established as a powerful tool for analyzing the fundamental properties of low-dimensional electron systems [2]. In phonon-drag studies an electric current is induced by (nonequilibrium) phonon pulses and is mapped as a function of phonon source position [3-6]. Because of the ability to isolate phonons of particular polarization and propagation directions this allows a direct insight into electron-phonon coupling mechanisms. The fact of the stronger and more anisotropic electron-phonon coupling in AlAs QWs leads to the conclusion that phonon-drag imaging should work even more successful with electrons in AlAs QWs than in comparable GaAs structures.

Up to now, almost all theoretical and experimental work on AlAs QWs, including phonon drag imaging [7-9], has been concentrated on (001) oriented wells. Other facets of growth, such as (110), have not been studied extensively. However, in the case of (001) oriented QWs, anisotropy effects are difficult to observe without application of an additional (external) symmetry breaking strain in the sample plane. The reason is that for a narrow (001) AlAs QW the electrons occupy only the out-of-plane X-valley with an isotropic Fermi contour. For wide QWs the two in-plane oriented X-valleys with anisotropic effective masses are occupied, but due to the effective superposition of the two valleys with orthogonal longitudinal mass axes the anisotropy is hidden.

Recent band-structure calculations and mobility measurements [10] have been shown that the valley occupancy in (110) oriented wells is different from those oriented in [001] direction. For wide (110) AlAs QWs (well width > 5 nm) single (in-plane) valley occupation is realized with a strong anisotropy of the in-plane elements of the inverse effective mass tensor and a resulting strong anisotropy of mobility [10,11]. For narrow (110) AlAs QWs an occupation of the doubly degenerate out-of-plane X-bands is expected. However, in contrast to the (001) AlAs QWs, also in this case an in-plane anisotropy is anticipated, because the projection of the two valleys onto the 2D electron plane gives two ellipses with collinear mass axes.

In this work we will present a theoretical study of how the phonon drag patterns of (110) AlAs QWs change with valley occupancy (well thickness).

## 2. Model and theory

We consider a system where the electrons are confined to an AlAs QW with $Al_xGa_{1-x}As$ barriers, grown on a (110) GaAs substrate, as used in the experiments of [10]. Bulk AlAs has an indirect band gap with conduction band minima at the six equivalent Brillouin zone X-points. The constant energy surfaces consist of three (six half) anisotropic ellipsoids with transverse and longitudinal effective band masses of $m_t = 0.19$ $m_e$ and $m_l = 1.1$ $m_e$, respectively. The ellipsoids (valleys) are labeled by the directions of their major axes in the crystal coordinate system: $X_x$, $X_y$, and $X_z$ for the [100], [010], and [001] valleys, respectively. Electrons confined in an AlAs QW have reduced valley degeneracy from the bulk. Which of the X-valleys is occupied depends on the growth direction and the QW width due to a balance of strain and confinement. The reason of the (biaxial) strain is the lattice mismatch between the AlAs layer and the GaAs substrate. Recent band structure calculations have shown that for (110) $Al_{0.45}Ga_{0.55}As/AlAs/Al_{0.45}Ga_{0.55}As$ QW structures with large enough well thickness (> 5.3 nm) only the two (half) $X_z$ valleys are occupied [10]. The major axis of the $X_z$-valleys lies in the (110) plane leading to a strong anisotropic elliptic Fermi surface. The band bottom of both out-of-plane valleys ($X_x$ and $X_y$) is higher in energy. For a well width less than 5.3 nm the role reverses and only the out-of-plane valleys with major axes along ±[100] and ±[010] are occupied. Their projection onto the (110) plane leads to elliptic constant energy surfaces with collinear mass axes.

In a typical phonon-drag experiment (see figure 1), a small active AlAs QW structure on one side of the GaAs substrate is irradiated with nonequilibrium acoustic phonon pulses generated on the other side of the substrate. The phonons absorbed in the AlAs QW transfer their in-plane component of quasimomentum to the 2D electrons and the resulting phonon-drag current is mapped as a function of the phonon source position. Therefore the phonon-drag images can be interpreted as a convolution of phonon focusing in the substrate (described by the magnitude of the incoming phonon signal in the QW) and the probability that a current will be induced by the phonons in the 2D electron system.

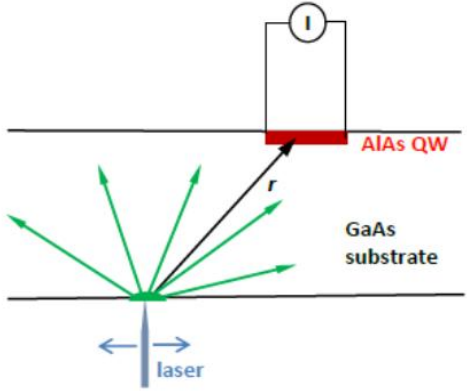

Figure 1. Schematic view of a typical phonon-drag imaging experiment.

The response of the 2D electrons due to the phonon pulse, i.e. the phonon drag current, can be naturally described by the motion of the center of mass of all 2D electrons. Using a method based on a Langevin equation approach to the quantum transport of interacting electrons, impurities and phonons, the time integrated current density in in-plane direction α is

$$\int dt\, j_\alpha(t,\mathbf{r}) = \frac{\mu_\alpha}{A} \sum_\nu \sum_{\mathbf{q},\lambda} q_\alpha \cdot \mathcal{N}_{\mathbf{q}\lambda}(\mathbf{r}) \cdot \left( \left| h_{\mathbf{q}\lambda}^{\nu(\mathrm{DP})} \right|^2 + \left| h_{\mathbf{q}\lambda}^{\nu(\mathrm{PE})} \right|^2 \right) \cdot \mathrm{Im}\left\{ \frac{|\mathcal{G}(q_{z'})|^2 \chi_0^\nu(q_{x'}, q_{y'}; \omega_{\mathbf{q}\lambda})}{\varepsilon(q_{x'}, q_{y'}; \omega_{\mathbf{q}\lambda})} \right\} \qquad (1)$$

(for details see [6,12] and the references therein). It is useful here to introduce a simulation coordinate system $(x', y', z')$, where the $z'$-axis is chosen along the [110] growth direction of the QW. The axes $x'$ and $y'$ are in the (110) plane, directed along the [-110] and the [001] direction, respectively. The sum on the right hand side of equation (1) is over all occupied valleys $X_\nu$ and over all phonon modes with phonon wavevector $\mathbf{q}$, polarization $\lambda$ and frequency $\omega_{\mathbf{q}\lambda}$. The electron mobility in direction $\alpha$ is denoted by $\mu_\alpha$ and $\mathcal{N}_{\mathbf{q}\lambda}(\mathbf{r})$ describes the time integrated number of (nonequilibrium) phonons with quasimomentum $\hbar\mathbf{q}$ and polarization $\lambda$ in the quasi-2D electron system of area $A$. The function $\mathcal{N}_{\mathbf{q}\lambda}(\mathbf{r})$ depends on the relative position $\mathbf{r}$ of phonon source and 2D electron devise, on the properties of the phonon source (like the wavevector and frequency dependence of the emitted nonequilibrium phonons) as well as on the phonon focusing in the GaAs substrate [4]. The overlap integral $\mathcal{G}(q_{z'})$ results from the confinement of the electrons in the $z'$-direction. For a QW of thickness $L_A$ we obtain

$$|\mathcal{G}(q_{z'})|^2 = \left( \frac{\pi^2}{\pi^2 - (q_{z'} L_A / 2)^2} \right)^2 \cdot \left( \frac{\sin(q_{z'} L_A / 2)}{q_{z'} L_A / 2} \right)^2 \qquad (2)$$

using the approximation of an infinite potential well. $\mathrm{Im}\,\chi_0^\nu(q_{x'}, q_{y'}; \omega_{\mathbf{q}\lambda})$ is the imaginary part of the dynamic susceptibility of a system of non-interacting 2D electrons in the $X_\nu$-valley. With the help of a Herring-Vogt like transformation one can rewrite the susceptibility of the anisotropic system in an isotropic form

$$\mathrm{Im}\,\chi_0^\nu(q_{x'}, q_{y'}; \omega) = -\frac{A m_0^\nu \kappa_F^\nu}{\pi \hbar^2 \xi^\nu} \mathrm{Re}\left\{ \sqrt{1 - \left( \frac{\xi^\nu}{2\kappa_F^\nu} - \frac{m_0^\nu \omega}{\hbar \kappa_F^\nu \xi^\nu} \right)^2} - \sqrt{1 - \left( \frac{\xi^\nu}{2\kappa_F^\nu} + \frac{m_0^\nu \omega}{\hbar \kappa_F^\nu \xi^\nu} \right)^2} \right\}, \qquad (3)$$

where $m_0^v = \sqrt{m_{x'}^v m_{y'}^v}$ is the density-of-state mass of the electrons in the X$_v$-valley. $\xi^v$ and $\kappa_F^v$ are defined by $\xi^v = \sqrt{q_{x'}^2(m_0^v/m_{x'}^v) + q_{y'}^2(m_0^v/m_{y'}^v)}$ and $\kappa_F^v = \sqrt{2m_0^v E_F^v/\hbar^2}$ with $E_F^v$ as Fermi energy of the electrons in the X$_v$-valley relative to the subband minimum. $(m_i^v)^{-1}$ denotes the diagonal element $(m^v)_{ii}^{-1}$ of the inverse effective mass tensor of the X$_v$-valley (with $i=x',y'$). So for the X$_z$-valley, for example, we get $m_{x'}^z = m_t$ and $m_{y'}^z = m_l$.

The electron-electron interaction we treat within the random-phase approximation which leads to screening of the electron-phonon coupling described by the dielectric function $\varepsilon(q_{x'}, q_{y'}; \omega_{q\lambda})$. For simplicity we use in our numerical calculations the static and zero electron temperature limit

$$\varepsilon(q_{x'}, q_{y'}; \omega_{q\lambda}) = \begin{cases} 1 + \dfrac{q_s(q_\parallel)}{\xi^v} & \text{for } \xi^v \leq 2\kappa_F^v \\ 1 + \dfrac{q_s(q_\parallel)}{\xi^v}\left(1 - \sqrt{1 - \left(2\kappa_F^v/\xi^v\right)^2}\right) & \text{otherwise} \end{cases} \quad (4)$$

with the screening wavevector $q_s(q_\parallel) = g_s g_v e^2 m_0^v g(q_\parallel)/4\pi\varepsilon_0\varepsilon_s \hbar^2$, where $g_s = 2$ is the spin degeneracy, $g_v$ is the valley degeneracy and $q_\parallel \equiv \sqrt{q_{x'}^2 + q_{y'}^2}$. The function $g(q_\parallel)$ is the form factor for a QW of width $L_A$ [13]

$$g(q_\parallel) = \left(\frac{\pi^2}{\pi^2 + (q_\parallel L_A/2)^2}\right)^2 \left\{\frac{2(e^{-q_\parallel L_A} + q_\parallel L_A - 1)}{(q_\parallel L_A)^2} + \frac{q_\parallel L_A}{4\pi^2}\left(5 + 3\frac{(q_\parallel L_A)^2}{4\pi^2}\right)\right\}. \quad (5)$$

For the electron-phonon interaction in AlAs QWs we have to include both deformation potential (DP) coupling

$$|h_{q\lambda}^{v(DP)}|^2 = \frac{\hbar}{2\rho V \omega_{q\lambda}}\left(\sum_{j=x',y',z'}(\Theta_d + \delta_{vj}\Theta_u)(e_{q\lambda})_j q_j\right)^2, \quad (6)$$

which is strongly anisotropic and depends on the considered X$_v$-valley, and piezoelectric (PE) scattering

$$|h_{q,\lambda}^{v(PE)}|^2 = \left(\frac{2e\,e_{14}}{\varepsilon_0\varepsilon_s}\right)^2 \frac{\hbar}{2\rho V \omega_{q\lambda}}\left(\frac{q_{x'}q_{y'}(e_{q\lambda})_{z'} + q_{y'}q_{z'}(e_{q\lambda})_{x'} + q_{z'}q_{x'}(e_{q\lambda})_{y'}}{q^2}\right)^2. \quad (7)$$

Here $e_{q\lambda}$ denotes the polarization vector of a phonon with wavevector $q$ and branch $\lambda$. $\Theta_d$ and $\Theta_u$ are the corresponding deformation potentials at the X-conduction valleys, $e_{14}$ is the piezoelectric stress constant and $\varepsilon_s$ the dielectric constant of the background material.

## 3. Numerical results and discussion

Our numerical simulations include the phonon focusing in the GaAs substrate, the GaAs/AlAs interface (with exception of mode conversion), the acoustic anisotropy of the electron-phonon coupling in the AlAs layer, the conduction-band anisotropy of the electrons in the AlAs QW and the geometry of phonon source and detector.

Figures 2 and 3 show the calculated patterns of phonon-drag current induced in (110) AlAs QWs of different well width (and of different valley occupancy) by beams of monochromatic phonons of frequency 300 GHz. The density of the quasi-2D electrons is in all cases $1.65 \cdot 10^{15}$ m$^{-2}$. Each point of the 2D maps corresponds to a respective position of the phonon source. The phonon source is moved in the (110) plane on the bottom side of the substrate and the scan from left to right (in $y'$-direction) and from bottom to top (in $x'$-direction) corresponds to an angular range in phonon propagation of $-51°…+51°$. The center of the image corresponds to the $z'$-direction (= [110] direction) and the drag current is measured in $y'$-direction. Positive and negative signals are represented as dark and bright shades; an average grey tone corresponds to zero signal.

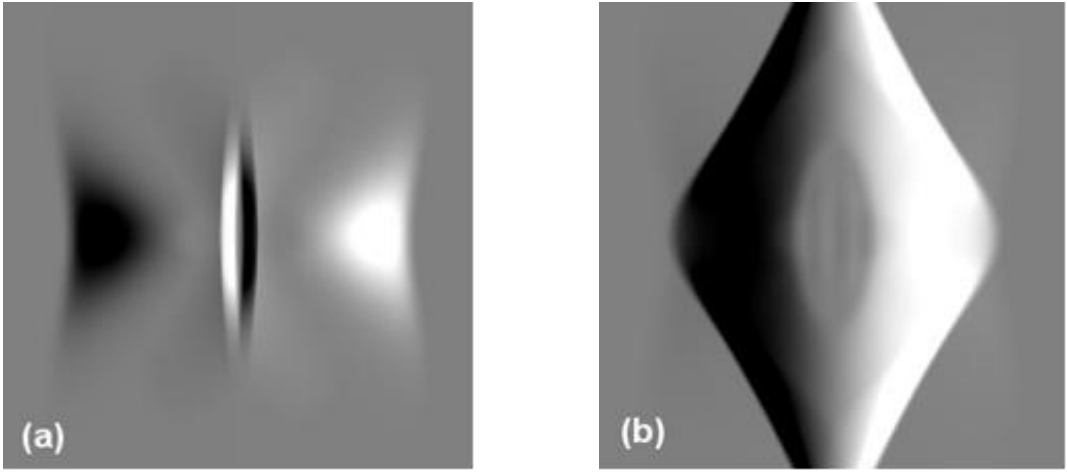

Figure 2: Calculated phonon-drag image of a 4 nm (110) AlAs QW ($X_x$- and $X_y$-valleys are occupied). In pattern (a) only PE electron-phonon coupling is taken into account, pattern (b) shows the contribution of DP interaction only.

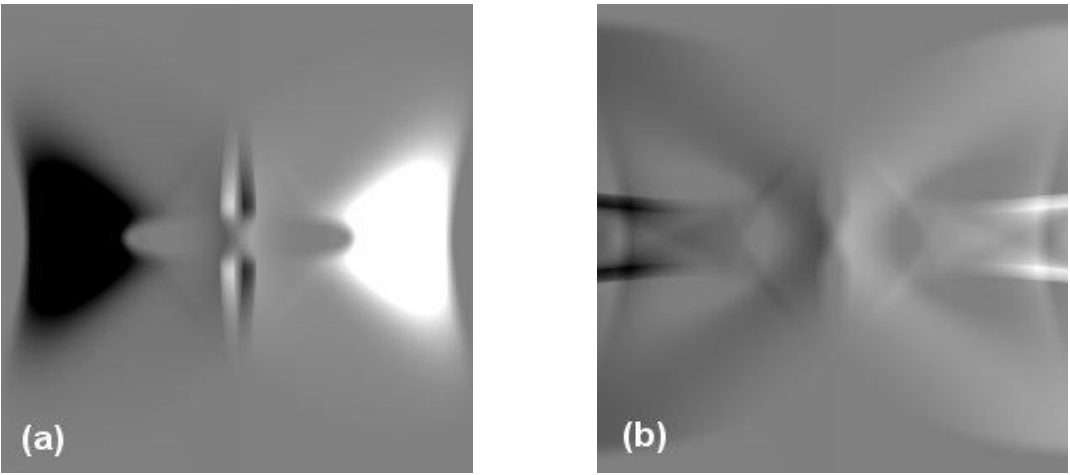

Figure 3. Calculated phonon-drag image of a 15 nm (110) AlAs QW ($X_z$-valley is occupied). In pattern (a) only PE electron-phonon coupling is taken into account, pattern (b) shows the contribution of DP interaction only.

The patterns presented in figures 2(a,b) are for a (110) AlAs QW of thickness $L_A = 4$ nm, where the $X_x$-and the $X_y$-valleys are occupied, while the patterns 3(a,b) correspond to the situation of wide wells ($L_A = 15$ nm), where the 2D electrons occupy only the $X_z$-valley. In our calculations we have used the following parameters for the deformation potential of the X-valleys: $\Theta_d = -1.1$ eV, $\Theta_u = 6.9$ eV and $\varepsilon_s = 10.1$ [14]. More difficult it is to find a reliable value for the piezoelectric coupling constant $e_{14}$. Several calculations of $e_{14}$ have been made in the past, and the results range from $-0.02$

Cm$^{-2}$ [15] to $-0.23$ Cm$^{-2}$ [16], whereas the latter value seems to be more accepted [14]. Since $e_{14}$ is squared, see equation (7), this uncertainty will be magnified. Therefore, we present the contributions of DP electron-phonon coupling and PE electron-phonon coupling separately. In the pattern 2(a) and 3(a) only the contribution of PE scattering is shown, while in 2(b) and 3(b) only DP interaction is included. If we take into account both interactions simultaneously, the PE coupling will clearly dominate for a value of $e_{14}$ close to $-0.23$ Cm$^{-2}$, whereas in the case of $e_{14} = -0.02$ Cm$^{-2}$ the main contribution would come from DP interaction. A comparison of our results with future experimental studies can therefore contribute to a more accurate determination of the coupling constants.

Due to the different occupied X-valleys the phonon-drag images for the 4 nm QW and the 15 nm QW differ significantly. Nevertheless, we can also observe common features for the narrow and wide QWs. In the patterns of PE coupling we find remarkably large amplitudes of the drag-current along two thin vertical stripes (along $x'$-direction), right and left of center. The reason of this effect is the acoustic anisotropy of the GaAs substrate, which causes a high flux of fast transversal acoustic (FTA) phonons into these directions, which can be validated from the corresponding quasimomentum focusing image (see figure 4). Additionally, we have a sign inversion along these FTA mode caustics. For larger in-plane components $q_{y'}$ of the phonon wavevector the current is mainly induced by longitudinal acoustic (LA) phonons in case of PE interaction and by slow transverse acoustic (STA) phonons in case of DP coupling. Such an unusually large contribution of DP coupled transverse phonons originates from the strong anisotropic $\Theta_u$-term in the DP electron-phonon matrix element (see equation (6)).

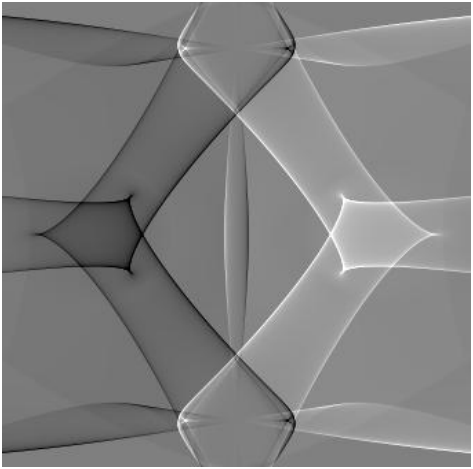

Figure 4. Calculated pattern of phonon focusing for a GaAs crystal. Measured is the [001] component of phonon quasimomentum ($\hbar q_{y'}$).

In the images of the 15nm AlAs QW (figures 3(a,b)) a strong suppression of the phonon-drag signal for large $x'$-components of the phonon wavevector is apparent. This is caused by the anisotropy of the effective mass tensor. Only the X$_z$-valley is occupied, where the minor and major axes of the highly anisotropic elliptic Fermi surface are characterized by $m_{x'}^z = 0.19\, m_e$ and $m_{y'}^z = 1.1\, m_e$, respectively. To study the influence of the anisotropic constant energy surfaces on the phonon-drag images in more detail, we contrast figures 3(a,b) with the corresponding images calculated for a (hypothetical) QW with an isotropic in-plane effective mass of $1.1\, m_e$ (figures 5(a,b)). The comparison clearly shows that the cutoff in the phonon absorption for large in-plane components of phonon wavevector is more effective in the direction of the smaller effective mass. Furthermore, we observe, according to equation (3), the expected increase of the drag-current amplitude with increasing density-of-state mass. The latter effect is verified by experiment [7] and is also confirmed by figure 6, where an isotropic in-plane effective mass of $0.19\, m_e$ was assumed.

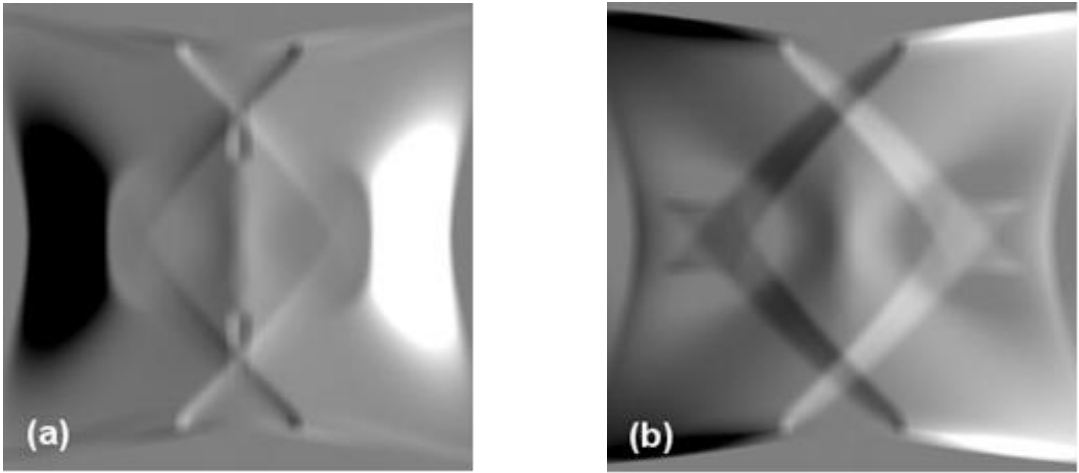

Figure 5. Calculated phonon-drag images of a hypothetical 15 nm (110) AlAs QW, where the occupation of an X-valley with isotropic in-plane effective mass of 1.1 $m_e$ is assumed. Pattern (a) is for PE electron-phonon coupling, pattern (b) for DP coupling.

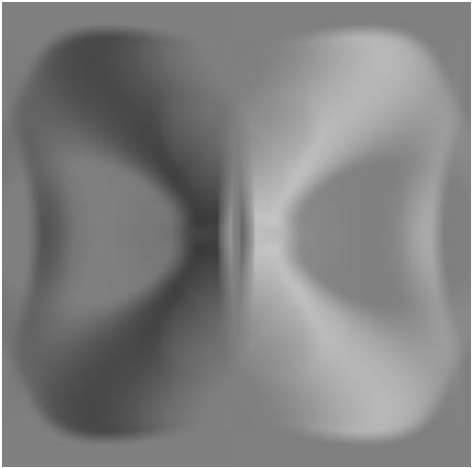

Figure 6. Calculated phonon-drag image (only DP coupling) of a hypothetical 15 nm (110) AlAs QW, where the occupation of an X-valley with isotropic in-plane effective mass of 0.19 $m_e$ is assumed. Due to the weaker phonon-drag current the signal is magnified by a factor 10 compared to figures 3 and 5.

## 4. Conclusions

We have presented a theoretical study of how phonon-drag patterns of (110) AlAs QWs change as a function of effective mass anisotropy and valley degeneracy. The influence of both the anisotropy in the electron dispersion and the anisotropy in the electron-phonon matrix elements can be clearly seen in the phonon drag patterns. Therefore our numerical simulations show that it is possible to obtain by phonon-drag measurements direct information about the valley occupancy of the electrons in the AlAs QWs and the effective mass tensor. This complements the results which may be obtained from the temperature dependence of the amplitude of Shubnikov-de Haas resistance oscillations [17,18] and from mobility measurements [10]. Furthermore, a comparison of our results with future phonon-drag imaging experiments allows a determination of the ratio of piezoelectric and deformation potential electron-phonon coupling parameters. This could, for the first time, open the possibility of a direct verification of the magnitude of the piezoelectric coupling constant $e_{14}$ in AlAs.